\documentclass{article}

\usepackage[usenames, dvipsnames]{color}
\usepackage[dvipsnames]{xcolor}

\usepackage{courier}
\usepackage{amsmath}
\usepackage{amssymb}
\usepackage{amsfonts}
\usepackage{amsthm}
\usepackage{graphicx}
\usepackage{tikz}
\usepackage{mathtools}
\usepackage{appendix}
\usepackage{xcolor}
\definecolor{xLightGray}{gray}{0.9}

\newcommand{\deriv}[2] {\partial_{#2} {#1}}

\newcommand{\reconstruct}[1]{\widetilde{#1}}

\newcommand{\ltype}{\mathbf}

\usepackage{caption}
\usepackage{hyperref}

\usepackage{listings}
\title{Axes that matter: PCA with a difference}

\author{Brian Huge \and Antoine Savine}
\date{First version March 2, 2021\\
        \vspace{1mm}
      This version \today}

\begin{document}

\maketitle

\begin{abstract}
 We extend the scope of differential machine learning and introduce a new breed of supervised principal component analysis to reduce dimensionality of Derivatives problems. Applications include the specification and calibration of pricing models, the identification of regression features in least-square Monte-Carlo, and the pre-processing of simulated datasets for (differential) machine learning.
\end{abstract}

\section*{Differential Machine Learning}

Giles and Glasserman's Smoking Adjoints \cite{Smoking} introduced algorithmic adjoint differentiation (AAD) to the financial industry and applied it to the computation of pathwise differentials (gradients of final payoff wrt initial state and model parameters) in the context of Monte-Carlo (MC) simulations. Although the customary application is the real-time computation of risk reports\footnote{We use the term \emph{risk reports} to refer to the collection of the risk sensitivities of a Derivatives transaction or trading book wrt current market state variables, or, in more formal terms, the gradient of its value wrt the state vector.} the availability of pathwise differentials also has enabled new research into more effective pricing and risk management practices.

Fast risk reports are computed by averaging pathwise differentials across scenarios. We argued in \cite{DML} that pathwise differentials have a bigger story to tell. We demonstrated that supervised machine learning (ML) models such as artificial neural networks (ANN) learn pricing and risk much more effectively from pathwise differentials. Differential ML more generally encompasses applications of AAD pathwise differentials in all kinds of ML. In this article, we address the problem of \emph{dimension reduction} and develop a new breed of \emph{differential} principal component analysis (PCA) to effectively  encode the market state into  low-dimensional representations. In contrast to the customary average of risk reports,  differential PCA leverages the \emph{covariance} of pathwise differentials.

The article starts with a brief discussion of dimensionality in Derivatives finance, where we motivate dimension reduction in the general context of Derivatives modeling and for least-square Monte-Carlo in particular. We follow with a discussion of classic, unsupervised algorithms such as PCA, where we highlight why they are not appropriate for Derivatives risk management. We introduce \emph{risk} PCA, which, in principle, provides a safe and effective dimension reduction mechanism, but suffers from prohibitive computation cost; and \emph{differential} PCA, which leverages AAD pathwise differentials to offer identical guarantees without the computation cost. The article concludes with a brief discussion of differential autoencoders and differential regression, and how they relate to differential PCA and the differential ANN of \cite{DML}.

\section*{Dimension reduction}

Consider the market state $X \in \mathbb{R}^n$ on some exposure date $T$. This means that the $n$ entries of $X$ are $T$-measurable random variables, which fully describe the state of the market on date $T$. Dimension reduction consists of an encoder function $g$ from $\mathbb{R}{^n}$ to $\mathbb{R}{^p}$ with $p \leq n$ such that the encoding $L=g\left(X\right)$ provides a lower-dimensional representation of the state $X$, while correctly capturing the information relevant for a given application. In ML lingo, the $p$ components of the latent representation $L$ are called meaningful \emph{features} of the state $X$.

Dimension reduction underlies the design and calibration of Derivatives models, albeit often in an implicit manner. Derivatives models are essentially a probabilistic description of the evolution of the market state $X$ between the exposure date $T$ and some maturity date $T^*$, generally written under a risk-neutral measure with parameters calibrated to implied volatility surfaces. We call \emph{universal} models probabilistic descriptions of the \emph{entire} state $X$. They generally require a heavy development effort. Their implementation is often slow, and their parameters are typically roughly calibrated to market data by best fit. Universal models are useful for complex exotics and hybrids, aggregated netting sets, or trading books for regulatory computations such as XVA or CCR, or the simulation of training datasets for learning pricing and risk functions as discussed in \cite{DML}. 

In many standard situations, more effective lightweight models are implemented by focusing on a small number of features relevant for a given application. For example, a basket option  depends only on the underlying basket under a Gaussian assumption, irrespective of its constituents. This remains approximately true under Black and Scholes, or other reasonably behaved local volatility models. In the absence of stochastic volatility, a basket option is essentially a one-factor product, so related models focus on a correct representation of the underlying basket  rather than single stocks. 

Similarly, a  European swaption depends only  on its underlying swap (and its volatility \footnote{For the purpose of this illustrative discussion, we only consider features of the \emph{underlying} market, assuming deterministic volatility. In the context of stochastic volatility, features of the volatility surface also affect transactions and are identified in the same manner.}), irrespective of the rest of the yield curve. Industry standard swaption models such as Black, Heston, or SABR, effectively model the underlying swap rate with increasingly complex dynamics, in contrast to term-structure models of the entire yield curve, suitable for more exotic instruments.

To identify the correct features is rather trivial with simple products and restrictive assumptions, but it is much harder in more complicated situations. And the stakes are considerable: to model irrelevant features is merely wasteful of computation resources, but to ignore relevant features may lead to incorrect prices, ineffective risk management, and wrong conclusions. For example, it was customary in the late 1990s to calibrate Bermudan option models to only co-terminal European swaptions. A more careful analysis in \cite{ AndersenAndreasen2001} demonstrated the strong dependency of Bermudan options on short rates and showed that ignoring the behavior of short rates implied by caplet volatility led to incorrect prices. Claims were made in the literature of a positive dependency of Bermudan option prices on interest-rate correlation \cite{LSBERM}, whereas the contrary is generally true with correctly calibrated models. Bermudan options are fundamentally (at least) two-factor products (ignoring stochastic volatility), so models must correctly represent the joint dynamics of co-terminal and short rates, and, in particular, simultaneously calibrate to co-terminal swaptions and caplets. 

Feature selection, or, equivalently, dimension reduction, is generally performed manually, often implicitly, and prone to error, with considerable consequences for pricing and risk. The objective of this article is to show how to perform dimension reduction in an explicit, principled, automatic, and reliable manner.

\section*{Least-square Monte-Carlo}

Another corner where dimension reduction is necessary is the ubiquitous least-square Monte-Carlo (LSM) designed by Longstaff-Schwartz \cite{LSM} and Carriere \cite{Carriere}, initially for Bermudan options in the context of high-dimensional Libor market models (LMM) implemented with MC. LSM estimates the continuation value on exercise dates by linear regression of future cash-flows on basis functions of the current state, iterating over call dates in reverse chronological order. Linear regression on basis functions of the high-dimensional state $X$ is intractable. Many recent articles suggest replacing regression with ML models robust in high dimensions such as ANN. But practitioners are often reluctant to forfeit the lightweight development, analytic efficiency (at least in a low dimension), and reliability of linear regression. The alternative is dimension reduction.

Practical LSM implementations generally walk around the curse of dimensionality by performing regressions on a low number of relevant features $L_i = g_i\left(X\right), i \in \left[1,p \right]$ of the state $X \in \mathbb{R}^n$. The resulting performance  directly depends on the correct selection of the regression variables $L_i$, usually performed manually and hardcoded in the implementation. For Bermudan options, many implementations pick the long-term swap to maturity and the short-term swap to the next call date. This is a reasonable choice for standard Bermudan options in the standard LMM model, but it doesn't scale to more complicated models or products. With stochastic volatility, for example, the volatility state constitutes another necessary feature. And guessing regression variables for callable exotics such as CMS spread steepeners may be perilous. What we need is a general algorithm to reliably identify the relevant features given an arbitrary transaction and an arbitrary pricing model.

In addition to the continuation values of callable instruments, LSM has been widely applied to the estimation of future values of trading books or netting sets in the context of regulations such as XVA, CCR, FRTB, or SIMM-MVA, where the repeated computation of prices and risks in multiple scenarios would otherwise consume unreasonable computing resources. Future values are usually computed in high-dimensional universal models, and their estimation is greatly facilitated by prior dimension reduction. We have empirically observed that dimension is often sufficiently reducible to enable lightweight linear regression. More robust approximators such as ANN also perform better in a lower dimension. Of course, handcrafted feature selection rules don't scale to trading books, so the need for a reliable algorithm is even more urgent in this context.

Dimension reduction algorithms are encapsulated in the encoding function $L=g\left(X\right)$. We will mainly explore linear encodings of the form $L=G X$, where $G\in\mathbb{R}^{p \times n}$ is a deterministic matrix.

Implementation details are covered in our GitHub repo\footnote{\url{https://github.com/differential-machine-learning/notebooks}}, along with Python code.

The notebooks \emph{DifferentialRegression} and \emph{DifferentialPCA} (best read in this order) discuss the implementation of the main algorithms, along with code and basic application examples. The more advanced \emph{Bermudan5F} notebook presents a practical application to Bermudan options, and reproduces the numerical examples that follow. All the notebooks are self-contained and run on Google Colab.  

\section*{Dimension reduction done wrong: classic PCA}

PCA is perhaps the simplest and certainly the most celebrated dimension reduction algorithm. It has been extensively applied in many scientific fields, including finance, in the literature and on trading desks, although this is, in our opinion, an incorrect tool to use for Derivatives. To better understand why, let us briefly recall the mechanics of PCA.

Consider an axis in $\mathbb{R}^n$ specified by the unit vector $u$. Define the \emph{variation} of $X$ along $u$ by the $\ltype L^2$ magnitude of the coordinates of $X$ on $u$:
$$
var_X\left(u\right) = E\left[\left(X \cdot u\right)^2 \right]
$$
A collection of $n$ mutually orthogonal unit vectors $u_1, ..., u_n$ constitutes an orthonormal basis of $\mathbb{R}^n$ so that the coordinates of $X$ can be rewritten in terms of this basis:
$$
    X = \sum_{i=1}^n \left(X \cdot u_i\right) u_i 
$$
PCA identifies a particular basis $\left(u_i\right)$ where the coordinates of $X$ are mutually orthogonal, measures the variation of $X$ along the axes $u_i$, and performs dimension reduction by truncation of the coordinates with immaterial variation. It is easily shown that an appropriate basis is given by the normalized eigenvectors of the covariance matrix of $X$, $C_X = E\left[X X^T\right]$, and that the corresponding eigenvalues measure the variation of the coordinates of $X$ in the eigenvector basis. PCA therefore reduces to eigenvalue decomposition:
$$
    C_X = P_X D_X P_X^T
$$
where $D_X$ is the diagonal matrix of eigenvalues and $P_X$ is the matrix of normalized eigenvectors in columns. The only data input is the covariance matrix $C_X$, usually estimated from a dataset of $m$ independent realizations of $X$. We do not discuss in this article  the subtleties of covariance, eigenvalue, and eigenvector estimation from finite datasets. This is covered in the statistics and ML literature. See \cite{MarcosEst} for an in-depth presentation along with robust estimation techniques. In what follows, we assume sufficiently large datasets for a correct estimation.\footnote{
The notations in this article are probabilistic. Here, $X$ denotes a random vector in dimension $n$, not a $m$-by-$n$ matrix with $m$ independent examples in rows and $n$ coordinates in columns, as customary in the statistics and ML literature. It follows that $C_X = E\left[X X^T\right]$ denotes the true covariance of $X$, and not its estimate $\hat C_X = \frac 1m {\mathbf X^T \mathbf X}$ from an $m$-by-$n$ dataset in a matrix $\mathbf X$. In practice, of course, the estimate is used in place of the ground truth covariance.}

With $P_X$ and $D_X$ conveniently sorted by decreasing eigenvalues, the first $p$ columns of $P_X$ span the subspace of $\mathbb{R}^n$ with maximum variation, and the remaining $q=n-p$ columns span the \emph{error space} where the variation of $X$ is minimized to $\epsilon = \sum_{i=p+1}^n \left(D_X\right)_{ii}$. Depending on the application, PCA is applied with fixed dimension $p$ or  fixed tolerance $\epsilon$, the latter being the safer option. By truncating the $q$ least significant axes of variation, PCA projects $X$ onto the subspace spanned by the first $p$ eigenvectors, resulting in the low-dimensional encoding:
$$
    L = G X \text{ with } G = \reconstruct D_X^{-\frac{1}{2}} \reconstruct P_X^T 
$$
where $\reconstruct P_X$ is the $n \times p$ matrix of the first $p$ columns of $P_X$, and $\widetilde D_X$ is the upper-left $p \times p$ corner of $D_X$, and normalization by $\reconstruct D_X^{-\frac{1}{2}}$  is optional but often convenient. Finally, $X$ is best reconstructed from $L$ in the sense of $\ltype L^2$ by application of the pseudo-inverse operator:
$$
    \widetilde X = H L \text{ with } H = G^T \left(G G^T \right)^{-1} = \reconstruct P_X \reconstruct D_X^{\frac{1}{2}} 
$$
which re-expresses $L$ in terms of the standard basis. PCA guarantees that the magnitude $ e_X = E \left[\left( X - \widetilde X\right)^2 \right] $ of the reconstruction error is bounded by $\epsilon$. In fact, PCA is equivalently expressed as the error minimization problem: $\min_{G \in  \mathbb{R}^{p \times n}} e_X $. 

Our objection is that PCA solves the wrong objective: we don't care about reconstruction fidelity. A large reconstruction error is acceptable when  aligned with a direction of insignificant risk. Equivalently, variation is not the correct metric: a small variation of $X$ in a direction of major risk may significantly affect the transaction. We don't want to rank axes by variation; we want to rank them by \emph{relevance} for a given transaction or trading book.

Consider, for instance, an option on the spread of two strongly correlated assets with similar volatilities. Here, the state $X=\left(X_1, X_2\right)$ is the pair of asset prices. Variation is concentrated along the diagonal $X_1 = X_2$, which accounts for a major fraction of the variance and constitutes the principal PCA axis. The orthogonal anti-diagonal   $X_2 = -X_1$ has negligible variation and is typically truncated by PCA. But the value of the spread option precisely depends on $X_2-X_1$, as revealed by gradients proportional to $\left(-1,1\right)$, exactly under Gaussian assumptions and approximately otherwise. Despite little variation, the anti-diagonal state coordinate is all that matters for the spread option. By truncating it, PCA completely misses value and risk. What is safe to truncate is the diagonal. Despite the strong variation, diagonal coordinates are irrelevant for the spread option, which value is essentially constant across all states on the diagonal axis. This very simple example illustrates that PCA is unsafe\textemdash it misses the principal axis of value and risk along the anti-diagonal\textemdash and ineffective\textemdash it fails to truncate the irrelevant diagonal. 

PCA fails precisely because it is completely unsupervised. It works only with state variation, irrespective of cash-flows. Like other unsupervised algorithms, \emph{PCA is one-size-fits-all}. Truncation is the same for all Derivatives; hence, it cannot be safe or effective. We have seen with the spread option example that PCA may truncate principal axes of value and risk. Consider now two Derivatives books: one is an option on a basket of stocks, the other is a basket of single-stock options. Under the Gaussian assumption, the first one is a one-factor product, depending only on the initial basket price. The second one, however, is a multifactor book whose risk cannot be summarized in one linear feature of the state. One-size-fits-all solutions cannot reduce dimensions correctly. We want to identify the axes that matter most for a given schedule of cash-flows and truncate the orthogonal space with negligible relevance. \emph{We need supervision. }

\section*{Dimension reduction done right: risk PCA}
Let us then supervise PCA and introduce price and risk labels:
$$
    V = v\left(X\right) \in \mathbb{R} \text{ and } \Delta = \frac{\partial v\left(X\right)}{\partial X} \in \mathbb{R}^n
$$
for a given transaction or trading book. To start building intuition, consider again an option on a basket of stocks with fixed weights $u$. Under the Gaussian assumption (and approximately otherwise), its value is a (nonlinear) function of the initial price of the underlying basket, that is $ v\left(X\right) = f\left(u \cdot X\right) $. It follows that the weight vector  $u$ is proportional to risk sensitivities:
$$
    \Delta\left(X\right) = \frac{\partial v\left(X\right)}{\partial X} = f'\left(u \cdot X\right) u \text{ hence }  u = \frac{\Delta\left(X\right)}{\left| \Delta\left(X\right) \right| }
$$
so the only meaningful feature $u \cdot X$ of the state $X$ is immediately identified by a single risk report in some arbitrary state. Notice in this case that the stochastic vector $\Delta\left(X\right)$ has deterministic direction $u$ and random magnitude $f'\left(u \cdot X\right)$. 

Of course, to identify risk factors with a single risk report computed in one given state $X$, will not work in general. Consider, for example, a delta-hedged option. The risk report in the current state cannot even see  the underlying stock as a risk factor. It is clear that multiple risk reports in different states are generally necessary to reliably identify all the relevant risk factors. To simulate a large number of risk reports is often intractable. The cost of pricing a complex transaction or trading book in many different scenarios is prohibitive, even with risk sensitivities efficiently computed with AAD. We will resolve this matter next. For now, we continue building intuition to formalize a dimension reduction algorithm. 

To generalize the previous example, consider a book that depends on $p$ linearly independent combinations  $w_i \cdot X$ of the state  $X$ (it may help to think of a collection of options on different baskets with weights $w_i$). The dependency on $w_i$ is easily re-expressed in terms of orthogonal unit vectors $u_i$, e.g. by Gram-Schmidt decomposition, such that:
$$
    v\left(X\right) = f\left(u_1 \cdot X, ..., u_p \cdot X \right) \text{ hence } \Delta\left(X\right) = \sum_{i=1}^p  \left(\deriv{f}{i} \right) u_i 
$$
where $\deriv{f}{i}$ denotes the derivative of $f$ wrt  the $i$-th argument. We see that the axes $u_i$ form a basis of the space of risk reports $\Delta$, thereby spanning the same $p$-dimensional subspace as the $p$ normalized eigenvectors with non-zero eigenvalues of the covariance matrix $C_\Delta = E \left[\Delta \Delta^T \right]$ of risk reports, of dimension $n$, and rank $p$. It follows that the number $p$ and directions $u_i$ of the relevant features are identified by the eigenvalue decomposition of $C_\Delta$.

Conversely, consider a dataset of risk reports where the risk wrt state variable $X_i$ is consistently zero, that is, ${\partial v\left(X\right)}/{\partial  X_i}=0$. It is clear, then, that, irrespective of the variation of $X_i$, the $i$th coordinate of $X$ is irrelevant for this particular transaction and safely truncated, so states are represented by remaining $n-1$ coordinates, and dimension is reduced by $1$. More generally, when the magnitude of directional risk $E\left[\left(u \cdot \Delta\right)^2\right]$ along some axis $u$ is immaterial wrt some threshold $\epsilon$,  direction $u$ is irrelevant and safely truncated, so that states are represented by their coordinates in the $\left(n-1\right)$-dimensional subspace orthogonal to $u$, without materially affecting the transaction. Those \emph{irrelevant} axes live in the error space of the covariance matrix $C_\Delta$ of risk reports, spanned by its $q$ least significant eigenvectors with cumulative eigenvalues bounded by $\epsilon$.

Those considerations suggest performing dimension reduction by PCA on risk reports $\Delta$ in place of states $X$. More formally, we define the weak relevance of some axis $u$ by the $\ltype L^2$ magnitude of risk along this axis (strong relevance is defined in the next section):
$$rel\left(u\right) = E\left[\left(u \cdot \Delta\right)^2\right] $$   
In other terms, relevance is the variation of risk. Orthogonal axes of decreasing relevance are given by the eigenvectors $P_\Delta$ of $C_\Delta$, where relevance is measured by the corresponding eigenvalues $D_\Delta$. Dimension reduction is performed by truncating $P_\Delta$ into $\reconstruct P_\Delta$ by removal of the  last $q$ columns: \emph{risk PCA is PCA on risks}. 

The lower-dimensional encoding of state $X$ is given by $L=G X$ with encoder $G=\reconstruct P_\Delta^T$ (normalization is unnecessary here), decoder $H=\reconstruct P_\Delta$, and reconstruction $\reconstruct X = H L$. The state reconstruction error $e_X = E\left[\left|X - \reconstruct X \right|^2 \right]$ is unbounded, because reconstruction fidelity is not the objective here. What is bounded by $\epsilon$ is the risk approximation error: 
$$e_\Delta = E\left[\left|\Delta - \Pi \Delta \right|^2 \right] =  E\left[\left|\Sigma \Delta \right|^2 \right] $$ 
where we defined the projection operator $\Pi = H G$ and the error operator $\Sigma = I_n - \Pi$. 

Risk sensitivities are re-expressed in terms of derivatives wrt $L$, denoted $S$:
$$
S = \frac{\partial \reconstruct V}{\partial L} = H^T \Delta \in \mathbb{R}^p
$$ 
where it is easily seen that risks wrt features $L$ are indeed orthogonal:
$$
    E\left[ SS^T \right] =  H^T C_\Delta H = \reconstruct D_\Delta
$$

Risk PCA therefore provides a safe dimension reduction where the magnitude of truncated risk does not exceed a specified threshold $\epsilon$. Because it implicitly minimizes the risk approximation error $e_\Delta$, risk PCA is optimal in the sense that further (linear) dimension reduction cannot be performed without increasing the magnitude of risk errors above $\epsilon$. Similar to classic PCA, the implementation is remarkably lightweight and limited to the eigenvalue decomposition of $C_\Delta$. However, because the estimation of $C_\Delta$ requires a simulated dataset of risk reports with prohibitive computation cost, risk PCA doesn't provide a practical algorithm, but merely a framework to reason about dimension reduction. 

Before we turn to the practical matter of performing risk PCA without the expense of simulated risk reports, let us briefly discuss a very simple and useful way to discriminate among factors affecting the transaction in linear or nonlinear manner. We introduced risk PCA by eigenvalue decomposition of the \emph{non-central} covariance matrix $C_\Delta = E\left[\Delta \Delta^T \right]$, decomposing $\mathbb{R}^n$ into directions of orthogonal risk, with magnitude measured by eigenvalues. An alternative representation is given by the eigenvalue decomposition of the usual, central covariance matrix $cov\left(\Delta\right)=C_{\Delta - E\Delta}$. The subtraction of the average risk effectively sets the magnitude of constant risks to zero. Directions of constant risk, which are axes of linear risk, are therefore truncated and only axes of nonlinear risk are picked up. Central risk PCA sorts and truncates axes by \emph{nonlinear relevance}. 

To see this more clearly, consider a second-order approximation whereby the Hessian matrix of $v$ is a constant $\Omega$, and input data $X$ is normalized so $EX=0$ and $E\left[XX^T\right]=I_n$. Then, it is easily seen that $cov\left(\Delta \right)=\Omega^2$ so the eigenvectors of covariance coincide with those of the Hessian. This correspondence forms the basis of modern variants of gradient descent widely applied in ML, such as RMSProp or ADAM, where a second-order modification is applied in reference to a Hessian matrix estimated from the covariance of gradients on previous iterations. 

In general, we want to correctly identify all directions of risk with a non-central risk PCA. In specific applications, we may want to consider only  nonlinear risk axes with central risk PCA. Notice that one effect of central risk PCA is to remove first-order hedges from the picture. A delta-hedged option, for example, is analyzed and encoded in an identical manner as the naked option. It is often useful in practice to perform both flavors of risk PCA.

\section*{Practical dimension reduction: differential PCA}

In order to turn risk PCA into a practical algorithm, we must circumvent the prohibitive simulation of a dataset $\left(X, V, \Delta\right)$ of $m$ independent value and risk reports. We solved the problem in our article \cite{DML}, in the spirit of the original LSM paper,  by learning the pricing function $v$ from a dataset of cash-flow samples $Y$ in place of prices $V$ and pathwise differentials $Z = \partial Y / \partial X$ in place of risk reports $\Delta = \partial V / \partial X$. Each training example $X$ is labeled by one payoff $Y$ (sum of cash-flows paid after horizon $T$) independently sampled by simulation of one MC path under the risk-neutral measure of the pricing model conditional to market state $X$ at $T$. Payoffs are sampled for the computation cost of one MC path, a fraction of the cost of pricing, and pathwise differentials $Z$ are efficiently computed with AAD for a small additional cost. The entire dataset of $m$ examples $\left(X, Y, Z\right)$ is therefore generated for a computation cost similar to \emph{one} pricing by MC, orders of magnitude cheaper than a collection of risk reports.

Given that $V=E\left[Y | X \right]$ (ignoring discounting) and $\Delta=E\left[Z | X \right]$ (assuming appropriate smoothing of discontinuous cash-flows), $Y$ and $Z$ are unbiased (if noisy) estimates of, respectively, $V$ and $\Delta$. This observation authorizes training on the tractable dataset $\left(X, Y, Z\right)$ in place of the prohibitively expensive  $\left(X, V, \Delta\right)$, a result we formally demonstrated by showing that pricing functions learned by universal approximators from the two datasets by minimization of the mean-squared-error (MSE) both converge to the correct pricing function in the limit of infinitely large datasets and infinite capacity.

Can we similarly approximate risk PCA, aka eigenvalue decomposition of $cov\left(\Delta\right)$, by \emph{differential PCA}, defined as the eigenvalue decomposition of $cov\left(Z\right)$? At first sight, the answer would be \emph{no}, because the covariance matrices do not coincide in general. By the law of total covariance:
$$
    cov\left(Z\right) = cov\left(\Delta\right) + E\left[cov\left(Z\right|X)\right]
$$
the additional covariance picked up in the simulation of the payoff does not vanish in the limit of infinitely large datasets and generally affects the eigenvalues and eigenvectors. It turns out, however, that differential PCA \emph{is} a close approximation of risk PCA, and, perhaps more importantly, a conservative one, due to the following key property. Denote $\Pi=\reconstruct P_Z \reconstruct P_Z^T$ the projection operator defined by differential PCA and $\Sigma = I_n - \Pi$ the corresponding error operator, then, using Jensen's inequality:
$$
 E\left[ \left| \Sigma \Delta  \right|^2 \right]  
                            = E\left[ \left| E\left(\Sigma Z|X\right) \right|^2 \right] 
                            \leq E\left[  E\left(\left|\Sigma Z\right|^2|X\right)  \right] 
                            = E\left[\left|\Sigma Z\right|^2  \right] 
$$

The true risk magnitude  truncated by differential PCA is \emph{lower} than the magnitude of truncated differentials. By construction of differential PCA, $e_Z = E\left[\left|\Sigma Z \right|^2 \right] \leq \epsilon$; hence, the approximation error of true risk is also bounded by $\epsilon$. By triangular inequality, the risk approximation from differential PCA and a hypothetical risk PCA are both distant from the true risk by less than $\epsilon$, so they must be at a distance less than $2 \epsilon$ from each other. It follows that differential PCA is a conservative, close approximation of risk PCA.

Differential PCA minimizes the differential reconstruction error denotd $e_Z$ and defined by $e_Z = E\left[\left|Z - \reconstruct Z \right|^2\right]$ and decomposes the state space according to a \emph{stronger} measure of relevance:
$$Rel\left(u\right) = E\left[\left(u \cdot Z \right)^2\right] $$   
Strong irrelevance implies weak irrelevance (and weak relevance implies strong relevance), so strongly irrelevant axes truncated by differential PCA are also weakly irrelevant; hence, they would be truncated by a hypothetical risk PCA, too. Differential PCA is safer than risk PCA. Intuitively, it correctly reconstructs differentials path by path, whereas risk PCA  reconstructs them only on average. Conversely, differential PCA may consider relevant some features that risk PCA would otherwise truncate. We can always relax the threshold $\epsilon$ accordingly. The implementation of differential PCA is otherwise a carbon copy of risk PCA, and all the comments made previously for risk PCA apply in an identical manner. 

The computational complexity of differential PCA is dominated by the simulation of the dataset, the computation of the covariance matrix $C_Z$, and its eigenvalue decomposition. We have seen that the dataset is efficiently simulated with AAD at a cost similar to one pricing by MC. This also holds for multiple exposure dates, because we can reuse the same simulated cash-flows, aggregated in different ways. The estimation of $C_Z=E\left[ZZ^T\right]$ has complexity $O\left(n^2m\right)$ for $m$ examples on each exposure date. An HPC library such as Intel's MKL computes the covariance of 32,768 examples in dimension 1,024 in about 0.25 seconds on a midrange trader workstation, or 1 second in dimension 2,048. Eigenvalue decomposition of the $n$-by-$n$ covariance matrix has cubic complexity. MKL computes the eigenvalues and eigenvectors of a real symmetric $1024 \times 1024$ matrix in about 0.05 seconds or 0.4 seconds in dimension 2,048. The cumulative complexity of other computations is insignificant. Differential PCA  takes only a fraction of a second to process one horizon date, a time negligible by risk computation standards, irrespective of the size of the trading book and in dimensions up to several thousands. When in doubt, processing is easily executed on a GPU with a library such as TensorFlow with another order of magnitude speedup.

\begin{figure}[ht]
\caption*{Numerical example: five-factor Gaussian interest-rate model}
\label{bermpca}
\centering
\includegraphics[width=\textwidth]{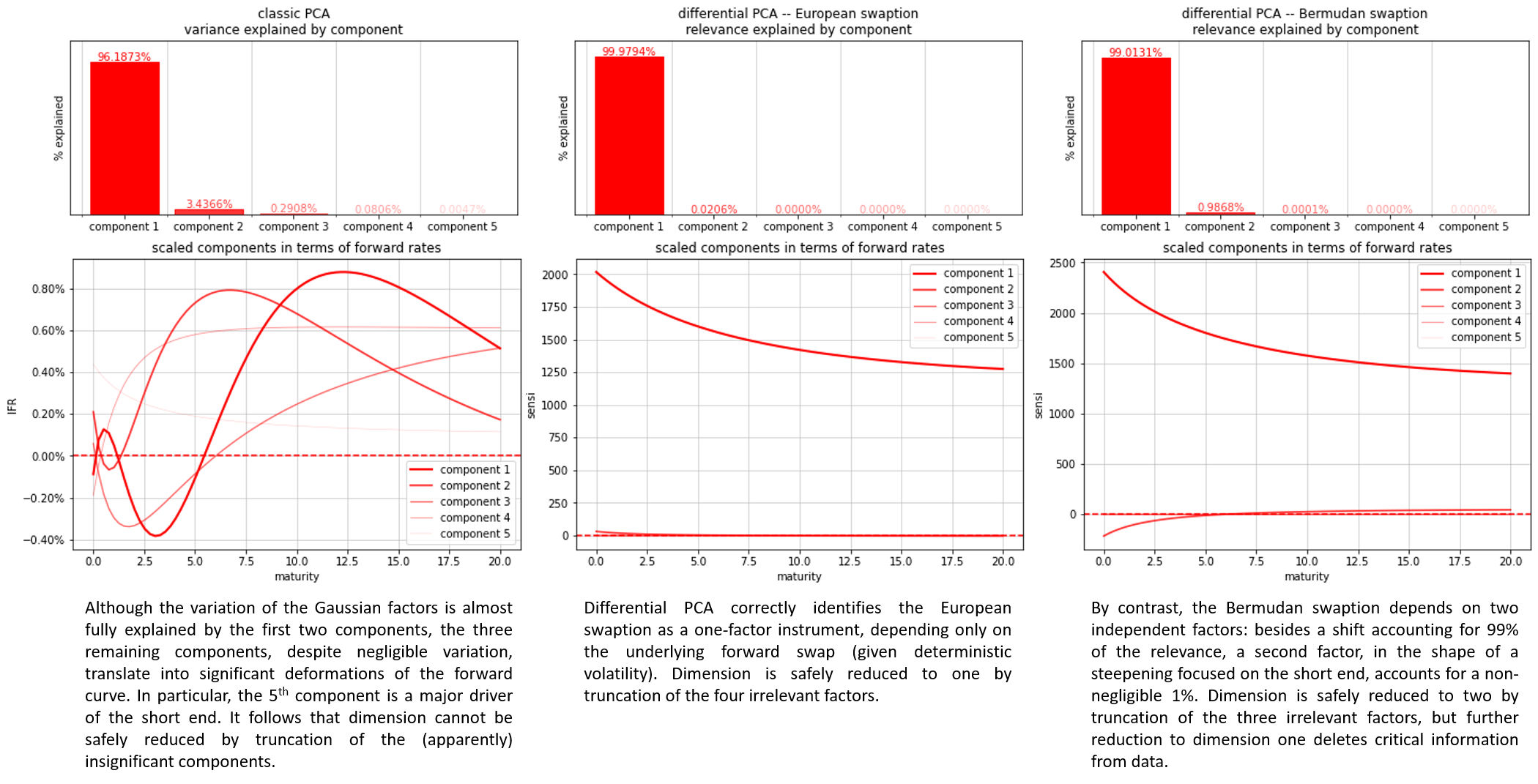}
Eigenvectors are expressed in terms of forward rates and normalized by root eigenvalue. See \emph{Bermudan5F.ipynb} on GitHub for detailed discussion and code.
\end{figure}

\section*{Nonlinear features and differential autoencoders}

Similar to its classic counterpart, differential PCA  identifies only linear features of the state of the form $L=G X$ (with reconstruction $\reconstruct X = H L$). Dimension may be further reduced by considering a wider class of nonlinear encodings of the form $L=g\left(X\right)$, with reconstruction $\reconstruct X=h\left(L\right)$. A now classic ML construct known as autoencoder (AE) effectively extends PCA to nonlinear encoding. The encoder $g: \mathbb{R}^n \to \mathbb{R}^p$ and decoder $h: \mathbb{R}^n \to \mathbb{R}^p$ are represented by two ANN, with, respectively, $n$ and $p$ units in their input layer and $p$ and $n$ units in their output layer, combining in one ANN with a characteristic bottleneck of size $p$ on the junction. The connection weights of the AE are trained by explicit minimization of the reconstruction error $e_X$, whose gradient wrt connection weights is efficiently computed by backpropagation on every iteration. AEs were also explored in finance, notably in \cite{CurveAE}, in the context of interest-rate curves. 

It is therefore tempting to extend differential PCA to differential AE, in an attempt to identify the most relevant nonlinear features of the state $X$ for a given transaction or trading book. But we already know the answer: the one nonlinear feature of the state that correctly encapsulates \emph{all} relevant information is the price $v\left(X\right)$. Learning relevant nonlinear features is identical to learning the pricing function $v$. We have discussed the pricing problem in depth in \cite{DML}, where we introduced differential ANN and demonstrated their effectiveness for learning pricing functions from differential datasets $\left(X, Y, Z\right)$. 

More specifically, the output layer of regression ANN is a linear combination of the nonlinear features encoded in the penultimate layer. The \emph{raison d'etre} of ANN is to learn relevant nonlinear features in their hidden layers. This is what distinguishes them from linear regressions, where regression features are fixed, and qualifies them as artificial intelligence (AI) constructs. This also explains why ANN are resilient in high dimensions, whereas linear regression is not. 

\emph{Differential autoeconders are implicitly embedded in differential ANN}. Although differential PCA identifies linear features of the state, of which the price is a nonlinear function, differential ANN identify nonlinear features of the state and combine them linearly into a price. The inspection of the features encoded in the penultimate layer provides insight into price formation and helps explain and interpret pricing by ML. By way of comparison, the price of European calls in \emph{all models} with a positive forward is decomposed (ignoring discounting) into:
$$
    c\left(F, t; K, T \right) = F Q^F\left(F_T>K\right) - K Q\left(F_T>K\right)
$$
the difference between two nonlinear features of the state with clear interpretation, where $Q$ is the risk-neutral probability measure and $Q^F$ is the spot-neutral probability measure.

\section*{Differential linear regression}
One benefit of differential PCA is to potentially reduce dimension sufficiently to enable linear regression. There is no guarantee that dimension is sufficiently reducible in a given context, but we have empirically observed that this is often the case. When linear regression is applicable by prior differential PCA or otherwise, its differential variant offers considerable benefits. Note that a training set for differential PCA already includes differential labels, and those same differential labels may be further leveraged to improve regression.

Recall that linear regression approximates $Y$ by a linear combination $\beta \cdot \phi\left(X\right)$ of $k$ basis functions $\phi\left(X\right) \in \mathbb R^k$ of $X$, most commonly monomials or basis spline functions. Because the number $k$ of basis functions coincides with the dimension of the weight vector $\beta$ and grows exponentially with dimension $n$, linear regression quickly becomes  intractable and vulnerable to overfitting when dimension grows. The weight vector is found by minimization of the $MSE = E\left[\left( Y - \beta \cdot \phi\left(X\right) \right)^2 \right]$ with a unique and analytic solution: 
$$
    \beta^* = C_{\phi}^{-1} C_{\phi,Y}
$$
where $C_{\phi} = E \left[\phi\left(X\right) \phi\left(X\right)^T \right]$ and $C_{\phi,Y} = E \left[\phi\left(X\right) Y \right]$ are estimated over a training set. Inversion is customarily stabilized with singular value decomposition (SVD) and further Tikhonov-regularized, that is, $\beta^* = \left(C_{\phi} + \lambda I_k\right)^{-1} C_{\phi Y}$ for some regularization strength $\lambda$.

Linear regression does not use differential labels $Z$. We argued in \cite{DML} that when those labels are available, as they are in finance with AAD, a much stronger form of regularization can be applied by minimizing a combination of value and derivative errors. In the case of linear regression, we have a mixed optimization objective:
$$
    \beta^* = argmin \left\{E\left[\left( Y - \beta \cdot \phi\left(X\right) \right)^2 \right] + \sum_{i=1}^n \lambda_i E\left[\left( Z_i - \beta \cdot \deriv{\phi}{i} \left(X\right) \right)^2 \right] \right\}
$$
where $\deriv{\phi}{i}\left( X \right)=\frac{\partial \phi\left(X\right)}{\partial X_i}\in\mathbb R^k$, and the relative weights $\lambda_i$ are commonly set to $E\left [ Y^2\right] / E\left [ Z_i^2\right]\in \mathbb R$.

Contrary to other forms of regularization, differential regularization does not introduce a bias and more generally massively improves the performance of ML models, as discussed in \cite{DML} in the context of ANN. The conclusions carry over to linear regression as a particular and simple case. In addition, the solution remains analytic and given by a modified normal equation, easily found by setting the gradient of the differential objective to zero:
$$
    \beta^* = \left( C_{\phi} + \sum_{i=1}^n \lambda_i C_{\deriv{\phi}{i}} \right)^{-1} \left( C_{\phi Y} + \sum_{i=1}^n \lambda_i C_{\deriv{\phi}{i},Z_i} \right)
$$
where
\begin{eqnarray*}
C_{\deriv{\phi}{i}} & = & E \left[\deriv{\phi}{i}\left(X\right) \deriv{\phi}{i}\left(X\right)^T \right]\in\mathbb R^{k\times k}\\
C_{\deriv{\phi}{i},Z_i} & = & E \left[\deriv{\phi}{i} \left(X\right) Z_i \right]\in\mathbb R^k
\end{eqnarray*}
are estimated over a training set. 

\begin{figure}[ht]
\caption*{Continuation value of the Bermudan option in the five-factor model, learned from 8,192 examples}
\centering
\includegraphics[width=\textwidth]{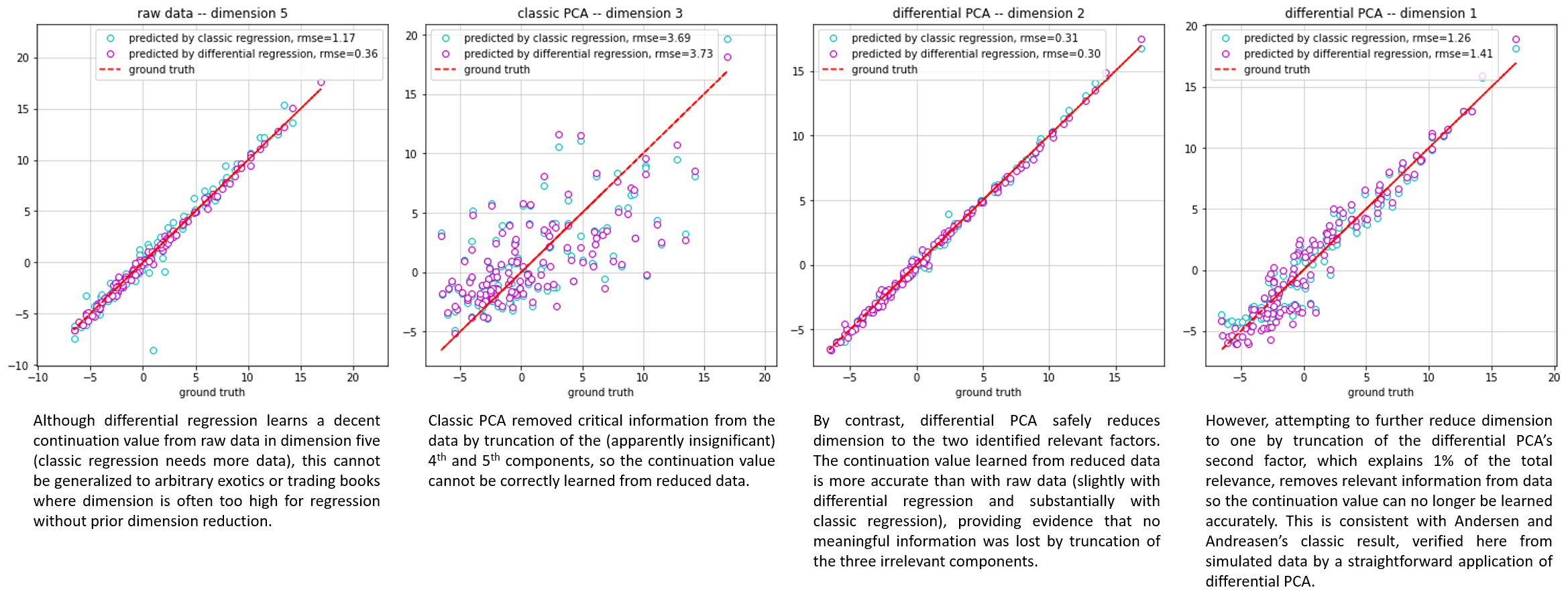}
Performance measured on 128 out-of-sample scenarios against ground truth. See \emph{Bermudan5F.ipynb} on GitHub for detailed discussion and code.
\end{figure}

\section*{Conclusion}

We extended the ideas of differential machine learning introduced in \cite{DML} to resolve the matter of dimension reduction, introducing the safe, effective, and efficient differential PCA. Similar to differential ANN or regression, differential PCA is based on AAD pathwise differentials. Unlike differential ANN or regression, differential PCA is not just a much more effective variant of PCA but also a very different algorithm. PCA cannot safely reduce dimension for a given collection of event-driven cash-flows, even in the limit of an infinite dataset. \emph{Differential} PCA does.  

The factors revealed by differential PCA help explain and manage the specific risks of a given trading book and calibrate pricing models in a reliable manner. In the context of least-square Monte-Carlo, differential PCA safely identifies a minimum number of regression variables, tying a longstanding loose end with application to Bermudan options and other callable exotics, as well as the valuation of Derivatives on ulterior dates. Differential PCA provides a very effective pre-processing of datasets for learning pricing and risk functions, often reducing dimension sufficiently to enable lightweight linear regression. Although more sophisticated ML models such as ANN are robust in a high dimension, they remain faster and more stable in a lower dimension. Differential PCA is lightweight, fast, and safe, hence, essentially zero-cost and best implemented as a systematic pre-processing step.

\bibliography{Bibliography}{}

\begin{thebibliography}{1}

\bibitem{AndersenAndreasen2001}
L.~Andersen and J.~Andreasen.
\newblock Factor dependence of bermudan swaptions: fact or fiction?
\newblock {\em Journal of Financial Economics}, 62(1):3--37, 2001.

\bibitem{Carriere}
J.~F. Carriere.
\newblock Valuation of the early-exercise price for options using simulations and nonparametric regression.
\newblock {\em Insurance: Mathematics and Economics}, 19(1):19--30, 1996.

\bibitem{Smoking}
M.~Giles and P.~Glasserman.
\newblock Smoking adjoints: Fast evaluation of greeks in monte carlo calculations.
\newblock {\em Risk}, 2006.

\bibitem{DML}
B.~Huge and A.~Savine.
\newblock Differential machine learning: the shape of things to come.
\newblock {\em Risk}, 2020.
\newblock Also available on SSRN and arXiv.

\bibitem{CurveAE}
A.~Kondratyev.
\newblock Curve dynamics with artificial neural networks.
\newblock {\em Risk}, June 2018.

\bibitem{LSBERM}
F.~A. Longstaff, P.~Santa-Clara, and E.~S. Schwartz.
\newblock Throwing away a billion dollars: the cost of suboptimal exercise strategies in the swaptions market.
\newblock {\em Journal of Financial Economics}, 62(1):39--66, 2001.

\bibitem{LSM}
F.~A. Longstaff and E.~S. Schwartz.
\newblock Valuing american options by simulation: A simple least-square approach.
\newblock {\em The Review of Financial Studies}, 14(1):113--147, 2001.

\bibitem{MarcosEst}
M.~{López de Prado}.
\newblock A robust estimator of the efficient frontier, 2019.
\newblock Available at SSRN.

\end{thebibliography}
\bibliographystyle{abbrv}

\end{document}